\documentclass[twocolumn,runningheads]{svjour2}

\smartqed  
\usepackage{graphicx}
\newcommand{\logg}{\mbox{$\log g$}}
\newcommand{\Teff}{\mbox{$T_\mathrm{eff}$}}

\newcommand{\msun}{\ensuremath{\, {\rm M}_\odot}}
 
\newcommand{\ion}[2]{\mbox{#1\,{\small #2}}}
\newcommand{\mdot}{\mbox{$\dot{M}$}}
\newcommand{\etal}{et\,al.\ }
\journalname{Astrophysics and Space Science}
\begin{document}

\title{Non-LTE modeling of supernova-fallback disks}


\author{Klaus Werner \and Thorsten Nagel \and Thomas Rauch}


\institute{K. Werner, T. Nagel, T. Rauch \at
              Institut f\"ur Astronomie und Astrophysik\\
              Universit\"at T\"ubingen, Germany \\
              \email{werner@astro.uni-tuebingen.de} 
}

\date{Received: date / Accepted: date}

\maketitle

\begin{abstract}
We present a first detailed spectrum synthesis calculation of a
supernova-fallback disk composed of iron. We assume a geometrically thin disk with a radial
structure described by the classical $\alpha$-disk model. The disk is represented by
concentric rings radiating as plane-parallel
slabs. The vertical structure and emission spectrum of each ring is computed in a
fully self-consistent manner by solving the structure equations simultaneously
with the radiation transfer equations under non-LTE conditions. We describe the
properties of a specific disk model and discuss various effects on the emergent
UV/optical spectrum.

We find that strong iron-line blanketing causes broad absorption features over
the whole spectral range. Limb darkening changes the spectral
distribution up to a factor of four depending on the inclination
angle. Consequently, such differences also occur between a blackbody
spectrum and our model. The overall spectral shape is independent of the exact
chemical composition as long as iron is the dominant species. A pure iron
composition cannot be distinguished from silicon-burning ash. Non-LTE effects
are small and restricted to few spectral features.

\end{abstract}

\begin{table*}[t]
\caption{Characteristics of the nine rings that compose the disk model. Surface
  mass density $\Sigma$ and emergent flux (expressed as \Teff) follow from the $\alpha$
  disk prescription. The Rosseland optical depth $\tau_{\rm
  Ross}$ at the disk midplane, the mass density $\rho$ and gravity \logg\ at optical depth unity
  follow from our computations of the detailed vertical ring structure. The last
  column denotes the fraction of the disk area that is made up by each ring
  model in order to compute disk-integrated spectra.}
\centering
\label{tab:rings} 
\begin{tabular}{cccccccc}
\hline\noalign{\smallskip}
Ring&$R$& $\Sigma$ & \Teff &$\log\tau_{\rm Ross}^{\rm midplane}$&$\log \rho(\tau_{\rm Ross}=1)$&$\log g(\tau_{\rm Ross}=1)$& \% area \\
    &[1000~km]&[g/cm$^2$]& [1000~K] &                       & [g/cm$^3$]& [cm/s$^2$]&   fraction \\
\tableheadseprule\noalign{\smallskip}
1   &2.0 &  2.9    & 305   &  3.1                          &   -6.1 &7.7   &0.0025\\
2   &2.5 &  2.8    & 258   &  3.2                          &   -6.2 &7.5   &0.01\\
3   &3.5 &  2.7    & 201   &  3.3                          &   -6.4 &7.2   &0.034\\
4   &6.0 &  2.6    & 135   &  3.5                          &   -6.0 &6.7   &0.084\\
5   &9.0 &  2.6    & 100   &  3.7                          &   -5.9 &6.4   &0.19\\
6   &14 &   2.5    &  72   &  3.3                          &   -6.4 &6.0   &0.62\\
7   &25 &   2.3    &  46   &  3.5                          &   -6.9 &5.5   &1.70\\
8   &40 &   2.2    &  33   &  3.9                          &   -7.1 &5.1   &33\\
9   &200&   1.9    &   9.8 &  3.8                          &   -7.6 &3.6   &64\\
\noalign{\smallskip}\hline
\end{tabular}
\end{table*}

\section{Introduction}

Anomalous X-ray pulsars (AXPs) are slowly rotating ($P_{\rm rot}=5-12$~s) young
($\leq 100\,000$~yr) isolated neutron stars. Their X-ray luminosities ($\approx
10^{36}$~erg/s) greatly exceed  the rates of rotational energy loss ($\approx
10^{33}$~erg/s). It is now generally believed that AXPs are magnetars with
magnetic field strengths greater than $10^{14}$~G and that their X-ray
luminosity is powered by magnetic energy (Woods \& Thompson 2006). As an alternative
explanation the X-ray emission was attributed to accretion from a disk that is
made up of supernova-fallback material (van Paradijs \etal 1995, Chatterjee
\etal 2000, Alpar 2001). The
fallback-disk model has difficulties to explain IR/optical emission properties
of AXPs.  When compared with disk models, the faint IR/optical flux suggests
that any disk around AXPs must be very compact (e.g.\ Perna \etal 2000, Israel
\etal 2004).

The discovery of optical pulsations in 4U\,0142+61 which have the same period
like the X-ray pulsations (Kern \& Martin 2002) appears to be a strong argument against
the disk model. It was argued that reprocessing of the pulsed NS X-ray emission
in the disk cannot explain the high optical pulsed fraction, because disk
radiation would be dominated by viscous dissipation and not by reprocessed NS
irradiation (Kern \& Martin 2002). 
Ertan \& Cheng (2004), on the other hand, showed that these optical pulsations can 
be explained either by the magnetar outer gap model or by the disk-star dynamo 
model. Therefore, the observation of optical pulsations is not an argument 
against the disk model.
A spectral break in the optical spectrum of
4U\,0142+61 was discovered by Hulleman \etal (2004). This was also taken as an argument
against the disk model because the authors do not expect such strong features
from a thermally emitting disk.  The recent discovery of mid-IR
emission from this AXP (Wang \etal 2006), however, has strongly rekindled the
interest in studies of fallback-disk emission properties. While this mid-IR
emission was attributed to a cool, passive (X-ray irradiated) dust debris disk
by Wang \etal (2006), it was shown by Ertan on this conference that it can
be explained with a model for an active, dissipating gas disk. If true, then the
disk emission properties allow to conclude on important quantities, e.g., the
magnetic field strength of the neutron star can be derived from the inner disk
radius.

Independent hints for the possible existence of fall\-back disks come from
pulsars with particular spin-down properties. For example, the discrepancy
between the characteristic age and the supernova age of the pulsar B1757-24  was
explained by the combined action of magnetic dipole radiation and accretion
torques (Marsden \etal 2001).  Even more, the presence of jets from pulsars such like
the Crab and Vela can possibly be explained by disk-wind outflow interacting
with and collimating the pulsar wind (Blackman \& Perna 2004).

A fallback-disk model was proposed in order to explain the X-ray enhancement
following a giant flare of the Soft Gamma Repeater SGR\,1900+14
(Ertan \& Alpar 2003). The X-ray light curve is interpreted in terms of the relaxation
of a fallback disk that has been pushed back by the gamma-ray flare. This model
can also explain the long-term X-ray and IR enhancement light curves of the AXP
1E\,2259+58 following a major bursting epoch (Ertan \etal 2006).

The presence of a fallback disk around the stellar remnant of SN~1987A has been
invoked in order to explain its observed lightcurve which deviates from the
theoretical one for pure radioactive decay (Meyer-Hofmeister 1992). From the
non-detection of any UV/optical point source in the supernova remnant, however,
tight constraints for the disk extension can be derived (Graves \etal 2005).

To our best knowledge, the emission from fallback disks in all studies was
hitherto modeled with blackbody spectra. In view of the importance of disk
models for the quantitative interpretation of observational data it is highly
desirable to construct more realistic models by detailed radiation-transfer
calculations.

\section{Radial disk structure}

For the modeling we employ our computer code AcDc (Nagel \etal 2004), that calculates
disk spectra under the following assumptions. The radial disk structure is
calculated assuming a stationary, Keplerian, geometrically thin $\alpha$-disk
(Shakura \& Sunyaev 1973). As pointed out by Menou \etal (2001), for a comparison with
observational data one probably has to use a more elaborate model, because near
the outer disk edge the viscous dissipation and hence the surface mass density
decline stronger with increasing radius than in an $\alpha$-disk. However, the
purpose of the present paper is to look for differential effects of various
assumptions. Qualitatively, these effects can be expected to be independent of
the detailed radial disk structure. In any case, it would be no problem to carry
out the computations presented here with different radial structures.

The $\alpha$-disk model is fixed by four global input parameters: Stellar mass
$M_\star$ and radius $R_\star$ of the accretor, mass accretion rate \mdot, and
the viscosity parameter $\alpha$. For numerical treatment the disk is divided
into a number of concentric rings. For each ring with radius $R$ our code
calculates the detailed vertical structure, assuming a plane-parallel radiating
slab.

In contrast to a (planar) stellar atmosphere, which is characterized by \Teff\
and \logg, a particular disk ring with radius $R$ is characterized by the
following two parameters, which follow from the global disk parameters
introduced above. The first parameter measures the dissipated and then radiated
energy. It can be expressed in terms of an effective temperature \Teff:
$$T_{\rm eff}^4(R)=[1-(R_\star/R)^{1/2}]\,3GM_\star\dot{M}/8\sigma\pi R^3.$$ The
second parameter is the half surface mass density $\Sigma$ of the disk ring: $$
\Sigma(R)=[1-(R_\star/R)^{1/2}]\, \dot{M}/3\pi \bar{w}.$$ $\sigma$ and $G$ are
the Stefan-Boltzmann and gravitational constants, respectively. $\bar{w}$ is the
depth mean of viscosity $w(z)$, where $z$ is the height above the disk
mid-plane. The viscosity is given by the standard $\alpha$-parametrization as a
function of the total (i.e.\ gas plus radiation) pressure, but numerous other
modified versions are used in the literature. We use a formulation involving the
Reynolds number $Re$, as proposed by Kriz \& Hubeny (1986). We chose
$Re=15\,000$ which corresponds to $\alpha\approx 0.01$.

\begin{figure}
\centering
  \includegraphics[width=\columnwidth]{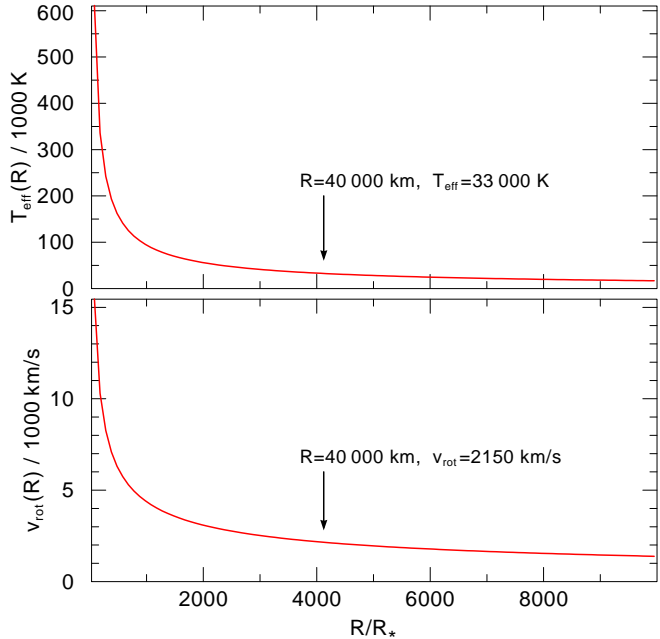}
\caption{Radial disk structure: Effective temperature (top panel) and Keplerian
  rotation velocity (bottom panel). Arrows mark a reference point at a distance
  of $R$=40\,000~km (about 4000 stellar radii R$_\star$) from the neutron star as discussed in
  the text.}
\label{fig:radial_structure}
\end{figure}

For the results presented here we selected the following input parameter
values. The neutron star mass is 1.4~\msun. The radii of the inner and outer
disk edges are 2000 and 200\,000~km, respectively. The disk is represented by
nine rings or, more precisely, by nine radial grid points. The radiation
integrated over the whole disk is then computed by assigning a weight to each
point's spectrum that resembles the area fraction that it represents.   The main
characteristics of the disk at the radial grid points are given in
Tab.~\ref{tab:rings}. The mass-accretion rate was set to $\dot{M}=3 \cdot
10^{-9}$~\msun/yr. Fig.~\ref{fig:radial_structure} shows the radial run of
\Teff. We also display the Keplerian rotation velocity for the later discussion
of our results. The radial distance from the neutron star is expressed in units
of the NS radius which is set to $R_\star$=9.7~km. But note from the above
equations that the disk model is essentially independent of the stellar radius
for large distances from the neutron star.

While $\Sigma(R)$ and \Teff$(R)$ in columns 3 and 4 of Tab.~\ref{tab:rings} follow
from the $\alpha$-disk assumption, the quantities in the next three columns are
the result from our detailed vertical structure calculations described below. It
shows that the entire disk model is optically thick. The Rosseland optical
depth at the disk midplane $\tau_{\rm Ross}^{\rm midplane}$ is $>$1000 at all
radii. We also tabulate the mass density and the gravity at unity optical depth.
That demonstrates that the conditions in the line forming regions of the disk
resemble those in white dwarfs at the inner disk radii up to main sequence stars
at the outer disk radii. The strength of Stark line broadening therefore
strongly depends on the distance of the emitting region from the neutron star.

\section{Vertical disk structure}

The vertical structure of each disk ring is determined from the simultaneous
solution of the radiation transfer equations plus the structure equations. The
latter ones invoke radiative and hydrostatic equilibrium plus charge
conservation. The structure equations also consist of the non-LTE rate equations
for the atomic population densities. The solution of this set of highly
non-linear integro-differential equations is performed using the Accelerated
Lambda Iteration (ALI) technique (Werner \& Husfeld 1985, Werner 1986, Werner
\etal 2003).

The total observed disk spectrum, which depends on the inclination angle, is
finally obtained by intensity integration over all rings accounting for
rotational Doppler effects.

\subsection{Radiation transfer, hydrostatic and radiative equilibrium}

We consider the radiation transfer equation for the intensity $I_{\nu}$ at
frequency $\nu$:
$$
\mu\,\frac{\partial\,I_{\nu}(\mu,z)}{\partial\,z}\,=\,-\kappa_\nu(z)\,I_{\nu}(\mu,z)\,+\,\eta_{\nu}(z)
$$
with the opacity $\kappa_\nu$ and the emissivity $\eta_{\nu}$. $z$ measures the
geometrical height above the disk midplane and $\mu$ is the cosine of the
inclination angle $i$. The equation is solved using a short characteristics
method. Opacities and emissivities are computed using atomic population
densities that are obtained by solving the non-LTE rate equations. Our code allows
for the irradiation of the disk by the central source, however, the results
presented here are computed with zero incident intensity.

\begin{figure}
\centering
  \includegraphics[width=\columnwidth]{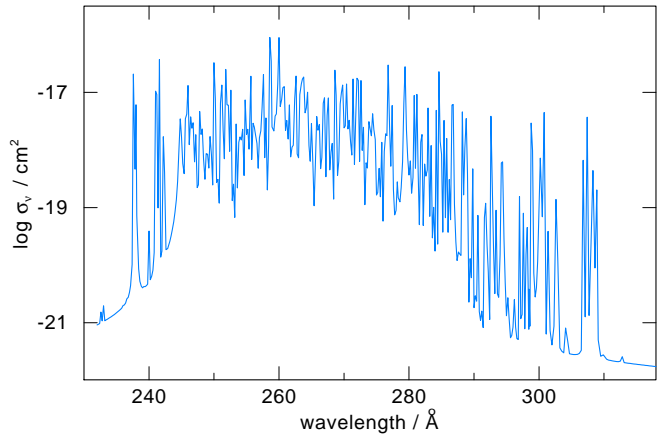}
\caption{Photon cross-section for a superline in the \ion{Fe}{IV} ion (between superlevels
  number 1 and 7).}
\label{fig:rbb}
\end{figure}

The radiation-transfer equations plus vertical structure equations are solved
like in the stellar atmosphere case, but accounting for two basic
differences. First, the gravity (entering the hydrostatic equation for the
total, i.e.\ gas plus radiation, pressure) is not constant with depth, but increases
with $z$. The gravity is the vertical component of the gravitational
acceleration exerted by the central object (self-gravitation of the disk is
negligible):
$$g=z\, GM_\star/R^3.$$ Second, the energy equation for radiative equilibrium
balances the dissipated mechanical energy and the net radiative losses:
$$ 9/4\,\, \rho w\, G\Sigma/R^3 = 4\pi \int_{0}^{\infty}(\eta_\nu-\kappa_\nu
J_\nu)d\nu ,$$ where $\rho$ and $J_\nu$ denote mass density and mean intensity, 
respectively. In the case of a stellar
atmosphere the left-hand side of this equation vanishes and we get the usual
radiative  equilibrium equation. The solution is obtained by a generalized
Uns\"old-Lucy scheme and yields the vertical temperature structure.

Having calculated the vertical structures and spectra of the individual
disk rings, the ring spectra are integrated to get the spectrum of the
whole accretion disk:
$$
I_\nu(i)\,=\,\cos(i)\int\limits_{R_{\rm{i}}}^{R_{\rm{o}}}\int\limits_{0}^{2\pi}I_\nu(i,\phi,r)\,r\,d\phi\,dr\,. 
$$
Here, $R_{\rm{i}}$ and $R_{\rm{o}}$ denote the inner and outer radius of
the disk, and $\phi$ is the azimuthal angle. At this stage, the Keplerian
rotation velocity v$_{\rm rot}$ is
taken into account by assigning a Doppler shift of $\Delta\nu=\frac{\nu}{c} {\rm
  v}_{\rm rot} \sin\phi \sin i$ to the intensity emerging from a specific azimuthal
ring sector.

\begin{table}[t]
\caption{Summary of non-LTE model atoms for silicon, sulfur, and iron. 
The numbers in brackets at the iron ions give the number of individual lines
summed up into superlines. Employed for a specific test run, the silicon and sulfur model atoms are tailored to the
conditions encountered in disk ring number~8.}
\centering
\label{tab:modelatoms} 
\begin{tabular}{llrrr}
\hline\noalign{\smallskip}
element & ion & NLTE levels & lines&\\
\tableheadseprule\noalign{\smallskip}
      Si & \mbox{\scriptsize III}   & 6  & 4  \\
         & \mbox{\scriptsize IV}    & 16 & 44  \\
         & \mbox{\scriptsize V}     & 1  & 0  \\
     \noalign{\smallskip}
      S  & \mbox{\scriptsize III}   & 1  & 0  \\
         & \mbox{\scriptsize IV}    & 6  & 4  \\
         & \mbox{\scriptsize V}     & 14 & 16 \\
         & \mbox{\scriptsize VI}    & 1  & 0  \\
      \noalign{\smallskip}
      Fe & \mbox{\scriptsize I}   & 7 & 25 & (141\,821)\\
         & \mbox{\scriptsize II}  & 7 & 25 & (218\,490)\\
         & \mbox{\scriptsize III} & 7 & 25 & (301\,981)\\
         & \mbox{\scriptsize IV}  & 7 & 25 & (1\,027\,793)\\
         & \mbox{\scriptsize  V}  & 7 & 25 & (793\,718)\\
         & \mbox{\scriptsize VI}  & 8 & 33 & (340\,132)\\
         & \mbox{\scriptsize VII} & 7 & 22 & (86\,504)\\
         & \mbox{\scriptsize VIII}& 7 & 27 & (8\,724)\\
         & \mbox{\scriptsize IX}  & 7 & 25 & (36\,843)\\
         & \mbox{\scriptsize X}   & 7 & 28 & (45\,229)\\
         & \mbox{\scriptsize XI}  & 1 &  0 & \\
\noalign{\smallskip}\hline
\end{tabular}
\end{table}

\subsection{Non-LTE rate equations}

For each atomic level $i$ the rate equation describes the equilibrium of rates
into and rates out of this level and, thus, determine the occupation numbers $n_i$:
$$
n_i\sum_{i\neq j}^{}P_{ij}-\sum_{j\neq i}^{}n_jP_{ji}=0 .
$$
The rate coefficients $P_{ij}$ have radiative and electron collisional components:
$P_{ij}=R_{ij}+C_{ij}$. The radiative downward rate for example is
given by:
$$
R_{ji}=\left(\frac{n_i}{n_j}\right)^{\star}4\pi\int_{0}^{\infty} 
\frac{\sigma_{ij}(\nu)}
{h\nu}\left(\frac{2h\nu^3}{c^2}+J_{\nu}\right)e^{-h\nu/kT}\,d\nu .
$$
$\sigma_{ij}(\nu)$ is the photon cross-section and
$({n_i}/{n_j})^{\star}$ is the Boltzmann LTE population ratio.

The blanketing by millions of lines from iron  arising from transitions between
some $10^5$ levels can only be attacked with the help of statistical methods
(Anderson 1989, Dreizler \& Werner 1993). At the outset, model atoms are constructed by
combining many thousands of levels into a relatively small number of
superlevels. The respective line transitions are grouped into superlines
connecting these superlevels. In this case, the population densities of the
superlevels are computed from the rate equations, in which the photon
cross-sections $\sigma_{ij}(\nu)$ in the radiative rates $R_{ij}$ do not contain
only a single line profile but all individual lines that are combined into a
superline. As an example we show such a cross-section in
Fig.\,\ref{fig:rbb}. The complete spectrum of our disk model
($\lambda$=4--300\,000~\AA) is sampled by 30\,700 frequency points.

\begin{figure}
\centering
\hspace{-3mm}
  \includegraphics[width=6.15cm,angle=90]{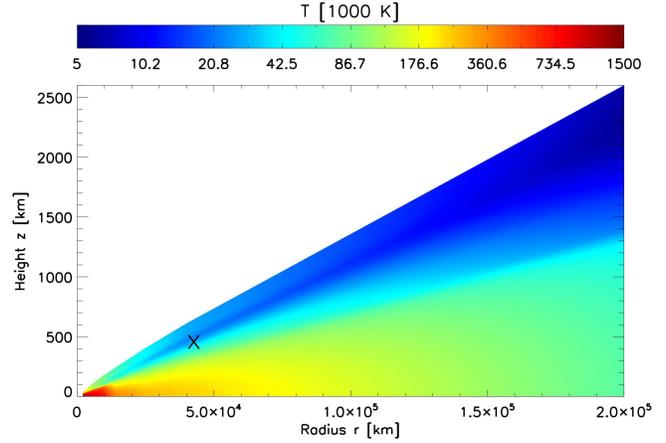}
\caption{This cut perpendicular to the midplane shows the temperature structure of the disk. Note that the vertical
  scale (height above the midplane) is expanded. The height-to-radius ratio is
  about 0.015. The cross marks the depth at $R$=40\,000~km where $\tau_{\rm Ross}=1$.}
\label{fig:Fe-Disctemperature}
\end{figure}

The model atoms that we have created for our disk calculations are summarized in
Tab.~\ref{tab:modelatoms}. Most important is the iron model atom. It comprises
the first eleven ionisation stages and a total number of more than 3 million
lines. Atomic data are taken from Kurucz (1991) and the Opacity and Iron
Projects (TIPTOPbase\footnote{ http://vizier.u-strasbg.fr/topbase/}).

\subsection{Disk composition}

The chemical composition of the supernova fallback material in the disk is not
exactly known. It depends on the amount of mass that goes into the disk. A disk
with a small mass (say $\leq 0.001$~\msun) will be composed of silicon-burning
ash (Menou \etal 2001). For simplicity, the results presented here are obtained by
assuming a pure-iron composition. It turns out that the emergent spectrum is
insensitive against the exact composition as long as iron is the dominant
species (Sect.~\ref{sect:abu}).  For a respective test run for one specific ring
we assumed a composition that represents a silicon-burning ash. It contains iron
(80\% mass fraction) as well as silicon and sulfur by 10\% each.

\subsection{Disk model properties}

Fig.~\ref{fig:Fe-Disctemperature} displays the temperature structure of the
disk. The temperature varies between 1.5 million K at the midplane at the inner
disk edge down to 6000~K in the upper layers at the outer disk edge. At all
radii the vertical run of the temperature decreases almost monotonously with
height above the midplane. A mild temperature reversal in the uppermost disk
layers occurs. This turns out to be a non-LTE effect, because the respective LTE
model has a strictly monotonous temperature run. We will discuss the
consequences of this effect in Sect.~\ref{sect:nlte}.

\begin{figure}
\centering
  \includegraphics[width=\columnwidth]{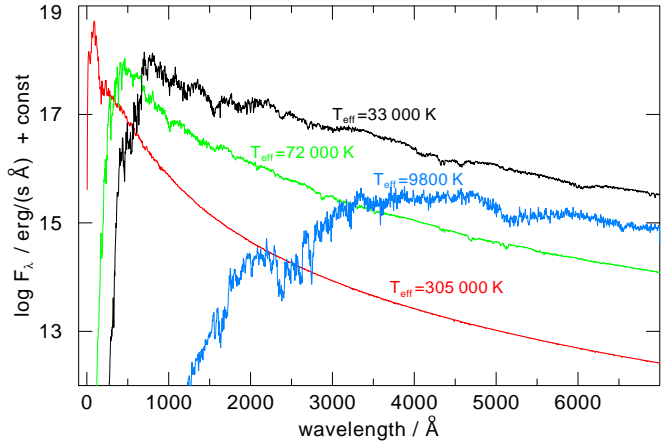}
\caption{Relative contribution of single disk rings to the total disk flux. For
  clarity, we only show the fluxes from rings number 1, 6, 8, and 9). The
  flux from ring~8 at $R$=40\,000~km with \Teff=33\,000~K dominates the
  total disk spectrum at UV/optical wavelengths.}
\label{fig:spectrum_contribution_of_single_rings}
\end{figure}

\begin{figure}
\centering
  \includegraphics[width=0.8\columnwidth]{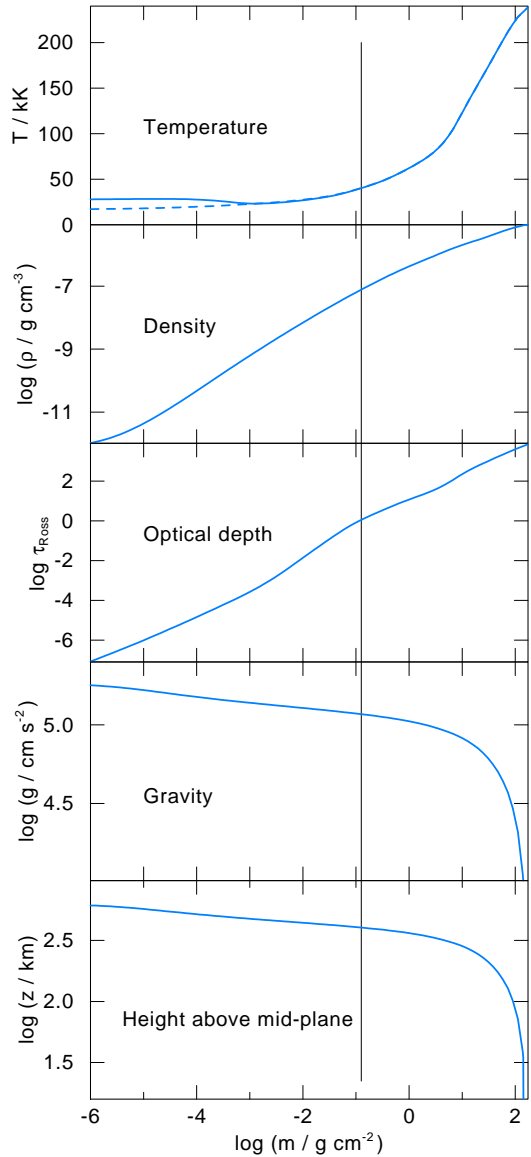}
\caption{Vertical stratification of the disk at $R$=40\,000~km (ring~8,
  \Teff=33\,000~K).  The quantities are plotted against the column mass density that is measured from the disk
  surface to the midplane (from left to right). The vertical line at $\log~m=-0.9$ denotes the depth where $\tau_{\rm
  Ross}=1$. The dashed curve in the top panel shows the temperature structure
  of a respective LTE model.}
\label{vertical_structure_ring8}
\end{figure}

Which disk regions contribute to the total disk spectrum and to what extent? In
Fig.~\ref{fig:spectrum_contribution_of_single_rings} we plot the emergent
astrophysical flux from the area of four disk rings (rings 1, 6, 8, and 9, see
Tab.~\ref{tab:rings}), i.e., the computed flux per cm$^2$ is weighted with the
ring area. The spectral flux distribution of the innermost ring with
\Teff=305\,000~K has its peak value in the soft X-ray region. The contribution
of this innermost region to the optical/UV spectrum is negligible. Cutting of
the disk at this inner radius ($R$=2000~km), therefore, is justified if this
spectral range is of interest. The disk region that is dominating the UV/optical
flux is represented by ring~8 with \Teff=33\,000~K. Its radius is 40\,000~km,
that is  about 4000 neutron star radii. Its spectrum is dominated by strong
blends of the numerous iron lines. Further out in the disk the effective
temperature decreases and the flux contribution to the UV/optical spectrum
decreases, too. Our outermost ring (number~9) has \Teff=9800~K, its flux maximum is
at $\lambda$=4000~\AA\ and it is fainter than the inner neighbor ring~8 over the
whole spectral range. Cutting off the disk at this outer radius
($R$=200\,000~km) therefore does not affect the UV/optical spectral region.

\section{Results}

Because ring~8 is a representative disk region that determines the UV/optical
spectrum, we will discuss its properties in more detail in the following three
subsections, addressing the effects of chemical composition, non-LTE physics,
and limb darkening on the spectrum.  The radius of this ring is marked by arrows
in Fig.~\ref{fig:radial_structure} and by a cross in Fig.~\ref{fig:Fe-Disctemperature}.

\begin{figure}
\centering
  \includegraphics[width=\columnwidth]{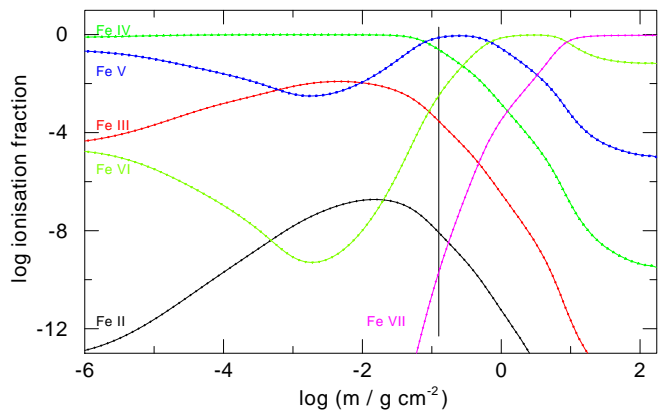}
\caption{Vertical iron ionisation structure at $R$=40\,000~km (ring~8,
  \Teff=33\,000~K). The vertical line at $\log~m=-0.9$ denotes the depth where $\tau_{\rm
  Ross}=1$. The dominant iron opacities in the line forming regions are from \ion{Fe}{III-VI}.}
\label{ionisation_structure_ring8}
\end{figure}

In Fig.~\ref{vertical_structure_ring8} we show the vertical structure of the
disk at this radius ($R$=40\,000~km). We plot the run of several quantities on a
column-mass scale, measured inward from the surface toward the midplane of the
disk. The vertical line at $\log m=-0.9$ marks the depth at which  $\tau_{\rm
Ross}=1$, i.e., the region of spectral line formation. The temperature shows a
strong increase towards the midplane and, as already mentioned, a mild
temperature reversal in the optically thin surface layers. Together with the
temperature run, the panels showing the mass density and gravity structure
indicate that their values in the line forming region are comparable to those
encountered in hot subdwarfs. The lowest panel shows that the disk height $H$ at
this distance from the NS is $\approx 600$~km, i.e.\ $R/H=0.015$.

In Fig.~\ref{ionisation_structure_ring8} we show the vertical ionisation
structure of iron in the disk at $R$=40\,000~km. The dominant ionisation stages
in the line forming regions are \ion{Fe}{III-VI}. At the midplane \ion{Fe}{VII}
is dominant. The temperature at any depth is so high that \ion{Fe}{I-II} do not
significantly contribute to the spectrum.

\subsection{Effects of chemical composition}
\label{sect:abu}

\begin{figure}
\centering
  \includegraphics[width=\columnwidth]{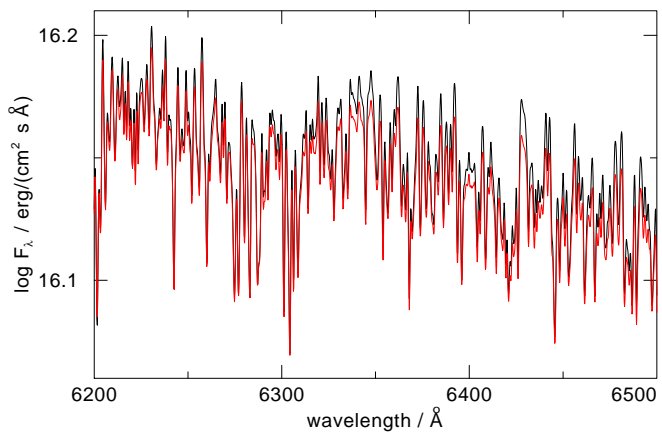}
\caption{Detail from the emergent disk spectrum at $R$=40\,000~km (ring~8,
  \Teff=33\,000~K). We compare a pure iron composition (black line) with a
  silicon-burning ash composition (red line). The differences are marginal.}
\label{spectrum_fe_vs_si_ash_ring8}
\end{figure}

In Fig.~\ref{spectrum_fe_vs_si_ash_ring8} we show the flux spectrum of ring~8 in
the wavelength range $\lambda$=6200--6500~\AA. It has been calculated for a pure
iron composition as well as for a Fe/Si/S=80/10/10 composition representing
silicon-burning ash. The difference between the spectra is very small, because
they are completely dominated by the extremely large number of iron lines. We
conclude that the exact disk composition is not affecting the spectrum as long
as iron is the dominant species.

While silicon and sulfur do not affect the overall spectrum by continuous
background opacities, line features can be seen in the computed spectra, e.g.\
the \ion{Si}{IV} resonance line in the UV. The line depth reaches about 50\% of
the continuum level but it would be difficult to detect even in medium
resolution spectra when the disk inclination is high and the spectra are Doppler
broadened by rotation.

\subsection{Significance of non-LTE effects }
\label{sect:nlte}

\begin{figure}
\centering
  \includegraphics[width=\columnwidth]{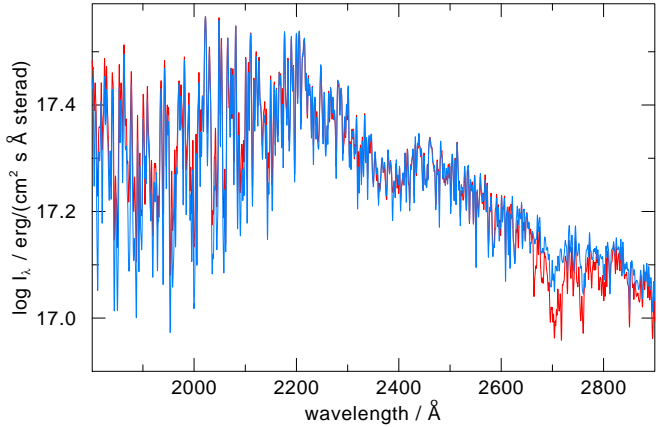}
\caption{Comparison of LTE and NLTE emergent disk intensities (red and blue lines,
  respectively) at $R$=40\,000~km (ring~8, $i=87^\circ$,
  \Teff=33\,000~K).}
\label{spectrum_lte_vs_nlte_ring8}
\end{figure}

For our particular disk model we expect that non-LTE effects are not very
large. This is because of the relatively high gravities in the line forming
regions, ranging between \logg=3.6 in the outermost ring with \Teff=9800~K and
\logg=7.7 in the innermost region with 305\,000~K. In
Fig.~\ref{spectrum_lte_vs_nlte_ring8} we  compare the spectra of an LTE and a
non-LTE model of disk ring~8 in the wavelength range $\lambda$=1800--2900~\AA,
where the largest deviations were found. Indeed, non-LTE physics affects only
narrow spectral regions. Only there, flux differences occur to an extent that
the equivalent width of line blends changes by a factor of two.  Accordingly,
the temperature structures of both models deviate only in the uppermost layers of
the disk (see top panel of Fig.~\ref{vertical_structure_ring8}) and, hence, only
strong spectral lines that are still optically thick can be affected.

\subsection{Limb darkening effects}

\begin{figure}
\centering
  \includegraphics[width=\columnwidth]{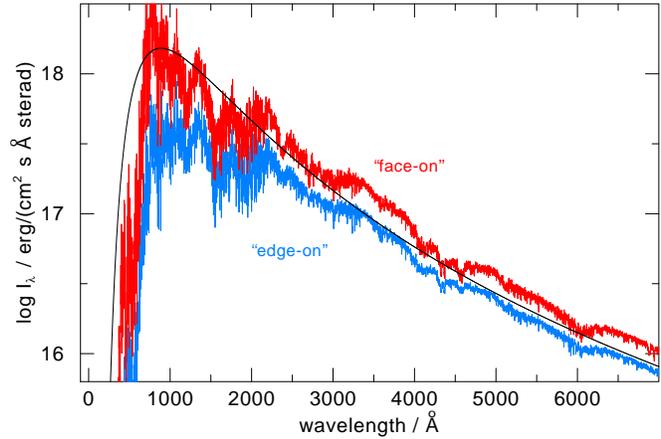}
\caption{Effect of limb darkening: Specific intensity of the disk at
  $R$=40\,000~km (ring~8) seen under two
  inclination angles, namely $87^\circ$ (i.e. almost edge-on, blue) and
  $18^\circ$ (i.e. almost face-on, red). For comparison we also show a
  blackbody spectrum with $T$=\Teff=33\,000~K.}
\label{spectrum_comparison_blackbody_ring8}
\end{figure}

Our model spectra show distinct limb-darkening effects. The situation is similar
to the stellar atmosphere case (center-to-limb variation of the specific
intensity). Looking face-on we see into deeper and hotter (and thus
``brighter'') layers of the disk when compared to a more edge-on view. In
Fig.~\ref{spectrum_comparison_blackbody_ring8} we compare the specific intensity
emitted by ring~8 (per unit area) for a high and a low inclination
angle. Overall, the ``edge-on'' spectrum is roughly a factor of two fainter than
the ``face-on'' spectrum in the optical region. The difference increases towards
the UV and amounts to a factor of about three. We conclude that limb darkening
effects are important when disk dimensions are to be estimated from magnitude
measurements.

It is also interesting to compare the intensities with a blackbody spectrum
(Fig.~\ref{spectrum_comparison_blackbody_ring8}). Depending on the wavelength
band, the blackbody over- or underestimates the ``real'' spectrum up to a factor
of two in the optical and a factor of four in the UV.

\subsection{Rotational broadening}

\begin{figure}
\centering
  \includegraphics[width=\columnwidth]{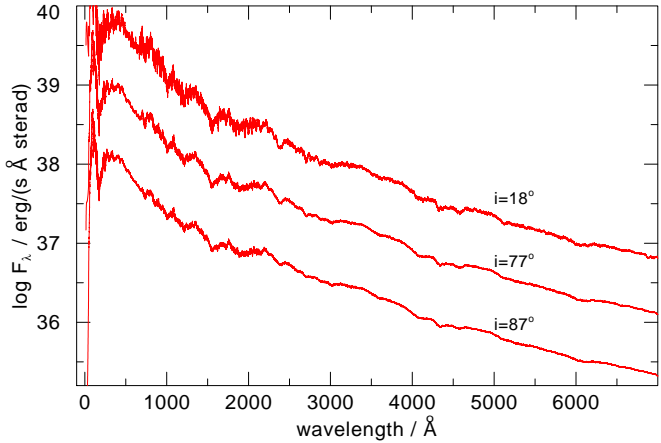}
\caption{Complete disk spectrum including Kepler rotation broadening, seen under three
  inclination angles. The broad iron-line blends are detectable even in the
  almost edge-on case.}
\label{spectrum_entire_disk_rotation}
\end{figure}

We have seen that the spectrum of an iron-dominated disk is characterized by
strong blends of a large number of lines. At some wavelengths broad spectral
features appear. It remains to be seen if Doppler effects from disk rotation
smears out these features or if they could still be detectable. From the bottom
panel of Fig.~\ref{fig:radial_structure} we see that rotational broadening
amounts to $\approx \sin i \cdot 2000$~km/s at $R$=40\,000~km, corresponding to
an orbital period of about two minutes. When seen almost edge-on, this
rotational velocity is equivalent to a Doppler broadening of about
$\Delta\lambda$=25~\AA\ at $\lambda$4000~\AA\ which clearly smears out any
individual line profiles. In Fig.~\ref{spectrum_entire_disk_rotation} we display
the rotationally broadened spectrum of the entire disk model, seen under three
different inclination angles. It is obvious that the broad line blends are so
prominent that they do not disappear even for an almost edge-on view of the disk.

Among the strongest features is a 200~\AA\ wide line blend at $\lambda$1500~\AA\
with an absorption depth of about 50\% relative to the continuum. Should the
disk be cooler, then disk regions with \Teff$\approx$9000~K could dominate the
optical emission and the disk spectrum might look more like that emitted by ring
9 in our model (Fig.~\ref{fig:spectrum_contribution_of_single_rings}). Strong
iron-line blanketing could cause a spectral break. This contrasts with a
statement in Hulleman \etal (2004), where the spectral break observed in the optical
energy distribution of the AXP 4U\,0142+61 is suggested as an argument against
disk emission.

\section{Summary and outlook}

We have computed a model for a supernova-fallback disk in order to study its
structure and optical/UV emission properties. We assumed an $\alpha$-disk for
the radial structure and performed detailed non-LTE radiation transfer
calculations for the vertical structure and spectrum synthesis. The input
parameters were: 
\\ \\ Neutron star mass:  1.4~\msun
\\ Inner and outer disk edge radii: $R$=2000 and 200\,000~km
\\ Mass-accretion rate: $\dot{M}=3\cdot
10^{-9}$~\msun/yr.  \\ \\ We have identified that the disk region in the
vicinity of $R$=40\,000~km is the main contributor to the total disk spectrum at
UV/optical wavelengths. We therefore investigated in some detail the disk
properties at this radius.

We summarize our results as follows:

\begin{itemize}

\item The overall disk spectrum is independent of the detailed chemical
composition as long as iron is the dominant species. In particular, a pure-iron
composition is spectroscopically indistinguishable from a silicon-burning ash
composition.

\item The overall disk spectrum is hardly influenced by non-LTE effects,
however, equivalent widths of individual line blends can change by a factor of
two.

\item Limb darkening affects the overall disk spectrum (in addition to the
geometric foreshortening factor $\cos i$). Depending on inclination and spectral
band, the disk intensity varies up to a factor of three.

\item Depending on the inclination, the disk flux can be a factor of two higher
or lower compared to a blackbody radiating disk.

\item Strong iron line blanketing causes broad ($>100$~\AA) spectral features
that could be detectable even from almost edge-on disks. Disks that are cooler
than our model (because of a lower mass-accretion rate) could even exhibit a
spectral break in the optical band due to massive line blanketing.

\end{itemize}

We stress that these results hold strictly only for our particular disk
model. In order to arrive at more general results a systematic parameter study
(disk extent, accretion rate) of the disk emission is necessary. Also, it needs
to be investigated in detail how fine the subdivision of the disk in a number of
rings is necessary in order to achieve a computed spectrum with a certain
accuracy.  In addition, deviations from the $\alpha$-disk model must be
studied. Another important point will be the inclusion of disk irradiation by
the X-ray emission from the neutron star. This will reveal the relative
importance of viscous dissipation and reprocessed irradiation that is discussed
in the context of simultaneous optical and X-ray pulsations in the AXP
4U\,0142+61. At the moment we do not dare to make any prediction how this
affects the results presented here.

The innermost disk ring has a very high effective temperature and its flux
distribution peaks in the soft X-ray band. It needs to be investigated
systematically under which conditions (inclination, inner disk radius, accretion
rate) the innermost disk regions can contribute to the thermal spectrum of the
magnetars.


\begin{acknowledgements}
T.R. was supported by the German Ministry of Economy and Technology through the
German Aerospace Center (DLR) under grant 50\,OR\,0201.
\end{acknowledgements}

\end{document}